\ifx\pdfvariable\undefined
\pdfoutput=1
\fi
\AtBeginDocument{\paperdetails{
  submitted=2017-04-07,
  published=2017-12-06,
  year=2018,
  volume=2,
  issue=2,
  articlenumber=3,
}}
\documentclass[english]{programming}
\usepackage[backend=biber]{biblatex}
\usepackage{listings}
\usepackage{enumitem}
\usepackage{graphicx}
\usepackage{tabularx, booktabs}
\usepackage{caption}
\newcolumntype{Y}{>{\centering\arraybackslash}X}

\usepackage{multicol}

\lstdefinelanguage{JavaScript}{
  keywords={typeof, new, true, false, catch, function, return, null, try, catch, switch, var, if, in, while, do, else, case, break, throw},
  keywordstyle=\color{blue},
  ndkeywords={class, export, boolean, implements, import, this, each, click, send},
  ndkeywordstyle=\color{blue},
  identifierstyle=\color{black},
  sensitive=false,
  comment=[l]{//},
  morecomment=[s]{/*}{*/},
  commentstyle=\color{purple}\ttfamily,
  stringstyle=\color{red},
  morestring=[b]',
  morestring=[b]"
}

\lstnewenvironment{listingcustom}[1][] 
 {\lstset{frame=tb, xleftmargin=.2\textwidth, xrightmargin=.2\textwidth, #1}}
 {}

\begin{document}

\title{Search-based Tier Assignment for Optimising Offline Availability in Multi-tier Web Applications}
\titlerunning{Search-based Tier Assignment for Optimising Offline Availability in Multi-tier Web Apps}
%\titlerunning{Preparing Articles for Programming} %optional, in case that the title is too long; the running title should fit into the top page column

\author[a]{Laure Philips}
\authorinfo[fig/laure]{is a PhD student at the Software Languages Lab, Vrije Universiteit Brussel in Belgium. Currently her main research area is web programming and more concretely tierless or multi-tier web programming. The first fruits of this research is Stip.js, a tier-splitting tool for tierless JavaScript code. Contact her at
\email{lphilips@vub.be}}
\author[a]{Joeri De Koster}
\authorinfo[fig/joeri]{ is Assistant Professor at the Software Languages Lab of the Vrije Universiteit Brussel in Belgium. He obtained the degree of Master of Science in Applied Sciences and Engineering: Computer Science in 2009 from the VUB and the degree of Doctor of Philosophy in Sciences: Computer Science in 2015 from the same institute. His current research interest lie in the design and implementation of programming languages for concurrent programming, big data processing, reactive programming and web development. Contact him at
\email{jdekoste@vub.be}}
\author[a]{Wolfgang De Meuter}
\authorinfo[fig/wolfgang]{is a professor at the Software Languages Lab, Vrije Universiteit Brussel in Belgium. Contact him at
\email{wdmeuter@vub.be}}
 \author[a]{Coen De Roover}
\authorinfo[fig/coen]{is a professor in Software Engineering at the Software Languages Lab of the Vrije Universiteit Brussel in Belgium. His research in program analysis and transformation techniques focuses on their design for modern programming paradigms (e.g., event-driven, actor-based, \ldots) as well as on their application to problems in software quality (e.g., bug detection, program renovation, change recommendation, \ldots).  Contact him at
 \email{cderoove@vub.be}}
\affiliation[a]{Vrije Universiteit Brussel, Brussels, Belgium}

 \authorrunning{L. Philips, J. De Koster, W. De Meuter and C. De Roover} % Optional, for long author lists

 % please provide 1--5 keywords

%%%%%%%%%%%%%%%%%%
%% These data MUST be filled for your submission. (see 5.3)
\paperdetails{
  %% perspective options are: art, sciencetheoretical, scienceempirical, engineering.
  %% Choose exactly the one that best describes this work. (see 2.1)
  perspective=art,
  %% State one or more areas, separated by a comma. (see 2.2)
  %% Please see list of areas in http://programming-journal.org/cfp/
  %% The list is open-ended, so use other areas if yours is/are not listed.
  area={Distributed systems programming},
  %% You may choose the license for your paper (see 3.)
  %% License options include: cc-by (default), cc-by-nc
  % license=cc-by,
}
%%%%%%%%%%%%%%%%%%

%%%%%%%%%%%%%%%%%%
%% These data are provided by the editors. May be left out on submission.
%\paperdetails{
%  submitted=2016-08-10,
%  published=2016-10-11,
%  year=2016,
%  volume=1,
%  issue=1,
%  articlenumber=1,
%}
%%%%%%%%%%%%%%%%%%

%%%%%%%%%%%%%%%%%%%%%%%%%%%%%
% Please go to https://dl.acm.org/ccs/ccs.cfm and generate your Classification
% System [view CCS TeX Code] stanz and copy _all of it_ to this place.
%% From HERE



% To HERE
%%%%%%%%%%%%%%%%%%%%%%%

\maketitle

% Please always include the abstract.
% The abstract MUST be written according to the directives stated in 
% http://programming-journal.org/submission/
% Failure to adhere to the abstract directives may result in the paper
% being returned to the authors.
\begin{abstract}
%Each submission must be accompanied by a plain-language abstract of up to 500 words that presents the key points in the paper in a manner understandable by experienced practitioners and researchers in %nearby disciplines. The abstract should avoid mathematical symbols whenever possible, and it must address the following:
%
%    Context: What is the broad context of the work? What is the importance of the general research area?
%    Inquiry: What problem or question does the paper address? How has this problem or question been addressed by others (if at all)?
%    Approach: What was done that unveiled new knowledge?
%    Knowledge: What new facts were uncovered? If the research was not results oriented, what new capabilities are enabled by the work?
%    Grounding: What argument, feasibility proof, artifacts, or results and evaluation support this work?
%    Importance: Why does this work matter?
%
% The absence of an abstract conforming to this specification is grounds for the rejection of the paper without review.

Web programmers are often faced with several challenges in the development process of modern, rich internet applications.
Technologies for the different tiers of the application have to be selected: a server-side language, a combination of JavaScript, HTML and CSS for the client, and a database technology. 
Meeting the expectations of contemporary web applications requires even more effort from the developer: many state of the art libraries must be mastered and glued together.
This leads to an impedance mismatch problem between the different technologies and it is up to the programmer to align them manually.
Multi-tier or tierless programming is a web programming paradigm that provides one language for the different tiers of the web application, allowing the programmer to focus on the actual program logic instead of the accidental complexity that comes from combining several technologies.
While current multi-tier approaches therefore relieve the burden of having to combine different technologies into one application, the distribution of the code is explicitly tied into the program.
Certain distribution decisions have an impact on crosscutting concerns such as information security or offline availability.
Moreover, adapting the programs such that the application complies better with these concerns often leads to code tangling, rendering the program more difficult to understand and maintain.
We introduce an approach to multi-tier programming where the multi-tier code is decoupled from the tier specification.
The developer implements the web application in terms of slices and an external specification that assigns the slices to tiers.
A recommender system completes the picture for those slices that do not have a fixed placement and proposes slice refinements as well.
This recommender system tries to optimise the tier specification with respect to one or more crosscutting concerns.
This is in contrast with current cutting edge solutions that hide distribution decisions from the programmer. 
In this paper we show that slices, together with a recommender system, enable the developer to experiment with different placements of slices, until the distribution of the code satisfies the programmer's needs. 
We present a search-based recommender system that maximises the offline availability of a web application and a concrete implementation of these concepts in the tier-splitting tool Stip.js.
 \end{abstract}

\section{Introduction}
\label{sec:introduction}
Application developers target the web platform more and more: not only is it supported on a plethora of devices, it is also an ideal stage for rich, interactive and collaborative applications. 
Such rich internet applications have a thick client that is more than just a static user interface.
The client integrates results from external services, e.g., a Twitter feed, and updates its UI reactively.
Moreover, not every click results in a roundtrip to the server and data can be stored locally, thus making (parts of) the application offline available as well.

Incorporating these characteristics requires the developer to master a multitude of technologies.
In a typical web application the developer has to select a technology stack for every tier: the client, server and database tier.
For example, Java or Php for the server, a query language for the database and a mingling of JavaScript, HTML and CSS for the client. 
Additionally, for the client tier, developers often resort to an extensive set of libraries for reactively updating the UI or doing complex rendering of data.
Manually aligning each of these different technologies frequently requires complex glue code that does not contribute to the essential complexity of the application. 
All of this additional accidental complexity lowers the maintainability of the resulting application.

Multi-tier or tierless programming effectively deals with these problems by enabling the development of multi-tier applications as a single artefact. 
Typical for this approach is that the same language can be used to develop all three tiers.
This enables a desktop-like style of development, because the accidental complexity resulting from the communication between the different tiers is hidden from the programmer.
Multi-tier programming was introduced over a decade ago and has since proven successful in solving several problems related to the web domain.

In this paper we distinguish three approaches to multi-tier programming: library-based, language-based and transformation-based.
Library-based approaches take a general-purpose language and provide an extensive set of libraries or packages that each handle a certain task related to the web domain.
Google Web Toolkit (GWT)~\cite{GWT:2010} is a multi-tier library for Java, and Meteor~\cite{meteor:2012}  for JavaScript.
While the library-based approach abstracts the domain-specific concepts into libraries, the language-based approach provides these concepts at the language level itself. 
A major advantage is that common tasks in the web domain that normally lead to complex code can now be abstracted away in new language features.
Examples include the Hop~\cite{hopjs:2016}, Links~\cite{links:2006}, Ur/Web ~\cite{ur/web:2015}  and ~\cite{opa:2013} multi-tier languages.
Transformation-based approaches also start from a general-purpose language, and the programmer can add developer hints via annotations.
These developer hints influence the transformation process that translated multi-tier to distributed code, and are related to the placement of code, remote communication, etc. The Volta tool \cite{volta:2008} takes programs in the Common Intermediate Language (CIL) of the .NET platform as input.

JavaScript has long become the standard for client-side browser development. 
Since the introduction of Node.js, the language can be used for server-side development as well. 
In previous work~\cite{Philips:2014, Philips:2015} we acknowledged this observation by introducing a transformation-based approach to multi-tier programming in JavaScript. 
Programmers can develop their applications as a single JavaScript artefact with tier-specific annotations that indicate what code needs to run on the client or server. 
An automatic tier splitter then calculates the two slices for the server and client tier and performs a code transformation to insert the necessary distributed logic.

By targeting JavaScript our tool can easily integrate with existing JavaScript libraries and tools (i.e., no typed interface or complex wrappers have to be provided).
Moreover, with this work we focus on a low-effort migration of existing JavaScript code where developers only have to add a few annotations in order to make their code tier-full.  

Up till now, multi-tier programming approaches have focussed on the elimination of the impedance mismatch between the different tiers in a multi-tier application. 
In order to do so, every approach has support for some tier-specific annotations in the source code that dictate what parts of the code have to run on each of the different tiers. 
For this paper, our hypothesis is that this explicit division of the source code into those tiers limits the configurability and flexibility of the resulting application.
More specifically, we argue that different crosscutting concerns such as performance, offline availability, security, etc. require different configurations of the same application.
For example, depending on the application, a developer might not know upfront what parts of the code need to run on the client, what parts on the server and what parts need to be replicated in order to maximise offline availability.
General-purpose data replication specifically, is a concern that has not yet been considered in the context of multi-tier programming.

In this paper we present an approach where applications are developed as a single artefact consisting of different \emph{slices}. Each slice represents part of the application that is written in a tier-agnostic way.
These slices serve as different building blocks for configuring an application. 
Such a configuration is specified in a placement specification where each of the slices are assigned to a specific tier. 
This assignment can be done manually or by means of a recommender system that automatically tries to calculate the optimal configuration in order to optimise a given metric for a certain crosscutting concern.
The assignment of different slices to different specific tiers in order to optimise this metric is a task that is often neither deterministic nor exact. For this reason our recommender system employs an evolutionary algorithm in order to calculate this assignment.

After assigning each slice to a tier the recommender system reports back to the programmer about the computed assignment of slices to specific tiers. On top of that, our recommender system will also suggest slice refinements that can potentially positively impact the result. The application developer can then split up or merge particular slices and insert or change specific annotations according to the suggested refinements. After this step, a next iteration of the recommender system can be started.

We argue that the combination of slices and the recommender system enables developers to focus first on the essential complexity of their applications. Crosscutting concerns can then later be tackled by changing the assignment of different sliced to specific tiers or by adding annotations in the code based on the suggestions of the recommender system.

This division of an application in tier-agnostic slices along with the recommender system are the main contributions of this work.
In this paper we introduce slice-based web programming, we first discuss our previous work on multi-tier programming.
We deliberate about possible approaches for the recommender system and give a concrete example of one that tries to optimise the offline availability of a web application. 
We give an evaluation by comparing different versions of the same application, that starts from a minimum of slices and incorporates the feedback from the system.
We measure how this has an impact on the offline availability, together with the granularity of the slices.
We implement the same application in a library-based and language-based multi-tier approach as well, and compare the runtime and code characteristics of the three versions.

\section{Slice-based web development}
\label{sec:slicebasedprogramming}
\subsection{Motivating example}
\label{sec:motivatingexample}
Throughout this paper we use a small example of a rich internet application, called Uni-corn. 
The idea is that PhD students can use this app to manage their \emph{``uni-versity''} career, by keeping track of tasks, meetings, teaching schedule, etc.
Through a built-in calendar view and charts page the student can monitor his or her progress.
This application has four main \emph{services}: viewing the calendar, viewing progress via the charts, viewing/adding/updating meetings and tasks.
To keep it simple we focus on a single server - single client application, meaning that there is no synchronisation between different clients.

The main focus of our motivating example is that the application has a high offline availability; meaning that the four services can still be used when no connection is at hand. 
For instance, entered data should not be lost when the connection drops, and ideally the newly added input should already be rendered. 
This way the programmer is not hampered by the lost connection and can keep using the application in the same way as before.

The application uses JavaScript libraries for representing and parsing dates, drawing the charts and rendering a calendar view. 
It is a single-page application, thus every view is already rendered when loading the application and no roundtrips to the server are made when switching between different views.
This also means that the user interface is updated reactively: any update on the data is immediately visible without a page refresh.
We published a full tutorial on how to write this application in our approach\footnote{\scriptsize{\url{http://lphilips.github.io/uni-corn.html} (accessed November 2017)}}.

\subsection{Slice-based web development with pre-determined tiers}
\label{sec:background}
In previous work~\cite{Philips:2014} we introduced a tier-splitting process for JavaScript that uses state of the art program analysis and program transformation technology. This section gives a brief overview of the contributions of that work.
The approach can be categorised as a transformation-based approach: a general-purpose language, namely JavaScript, is augmented with a number of light-weight developer hints to guide the splitting process. These developer hints are placed as annotations inside comment blocks. This means that any tool or library that works for JavaScript code is oblivious of these annotations and can still be used in conjunction with our approach. This means that developers can use and benefit from the entire JavaScript ecosystem.

In order for a developer to transform his tierless code into tierfull code different code blocks have to be annotated with either \textsc{@client} and \textsc{@server} annotations. The tier-splitting process will then automatically detect inter-tier communication and transform the source code into two program slices, one for the server and one for the client.
Annotations for remote communication, data sharing and failure handling are supported as well, an overview can found in Appendix~\ref{app:annotations}.

Our approach employs a state of the art program analysis tool for JavaScript in order to automatically infer where inter-tier communication needs to be inserted. 
This means that some of these annotations, related to inter-tier communication (e.g., \textsc{@remoteCall} and \textsc{@remoteProcedure}), can be omitted from the source code. This means that for those calls, a local call is indistinguishable from a remote call in the source code. Whether it is transformed is then solely determined by the placement of the caller and the callee.
When our static analysis loses precision and cannot determine whether a call should be remote or not this is reported and a manual annotation is required.

Based on the AST of the multi-tier code, we build the corresponding Program Dependence Graph (PDG) representation of the program.
It is a directed graph of which the nodes correspond to statements in the program and the edges represent control and data dependencies between the nodes.
PDG's are often used for program slicing~\cite{Weiser:1981}, a technique used for program comprehension and debugging.
Informally, a program slice is an executable subset of a program that has a direct or indirect effect on values computed at a certain location, or \emph{criterion}. 

\begin{figure}
\centering
 \includegraphics[scale=0.34]{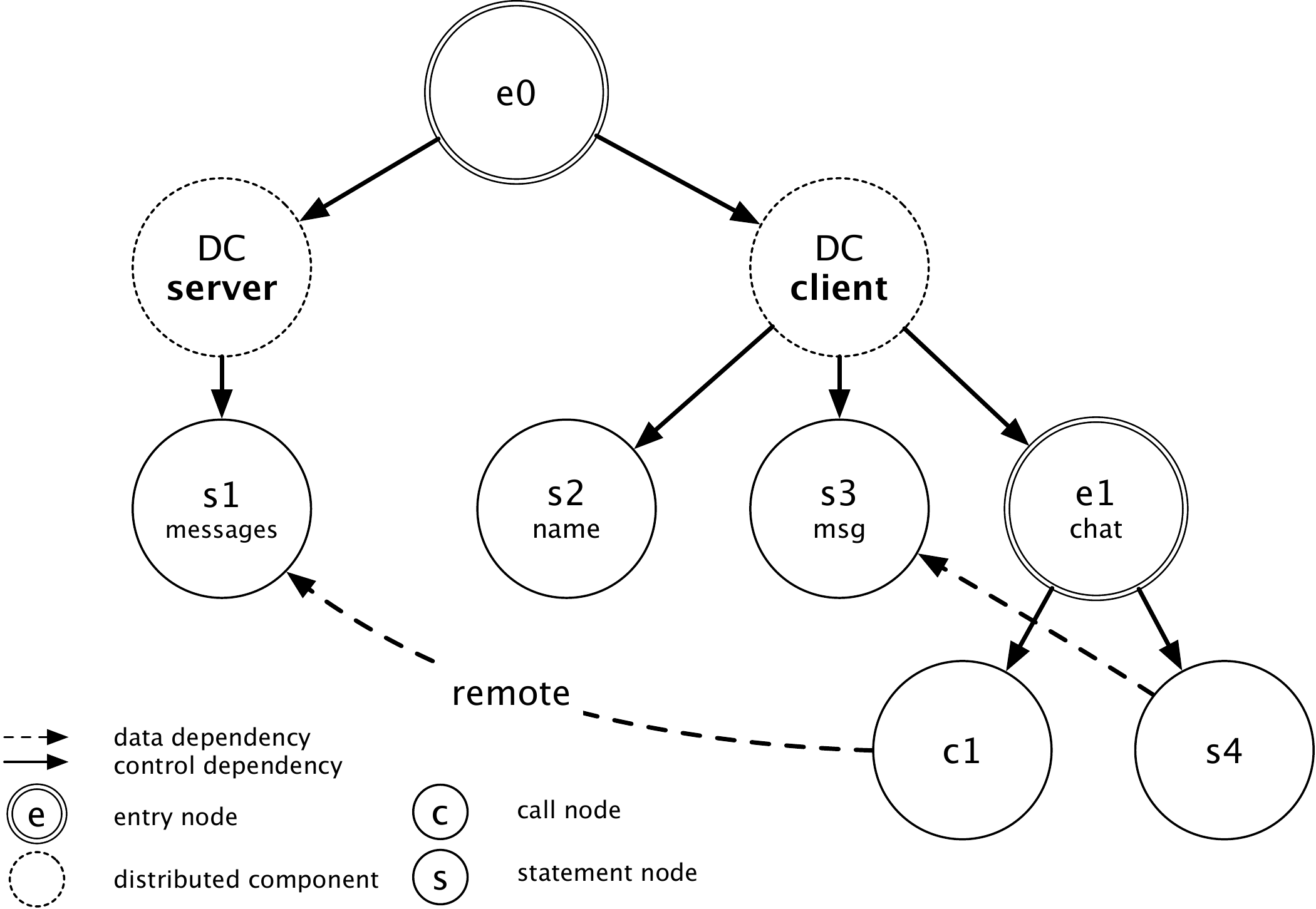}
\caption{Pogram Dependence Graph for Listing~\ref{lst:stipchat}}
\label{fig:pdg}
\end{figure}

We extended the PDG and program slicing algorithm with the notion of tiers, depicted by Distributed Component (i.e. client and server) nodes in the graph. 
Dependencies between nodes that belong to another distributed component are \emph{remote} dependencies: e.g., a remote data or call reference. 
Figure~\ref{fig:pdg} gives an example of a PDG for the code given in Listing~\ref{lst:stipchat}.
Our adapted slicing algorithm takes a distributed component as a slicing criterion and returns the subset of the code that is needed to execute that component, without including nodes that belong to another distributed component. 
Please note that not every node must belong to a distributed component, we call these shared nodes, because they end up in the program slices of those distributed components that use them.

After the program slicing step we end up with two program slices: one for the client tier and one for the server tier. 
The program transformation step now transforms these selected nodes to their distributed variants by injecting distributed communication into the multi-tier code. 
This means that local calls must become remote procedure calls and that local functions that are called by another tier must become remote functions.
The program transformation step calls on the information of the PDG to decide whether such transformations must take place.

In our prototype implementation of this tier-splitting process, called Stip.js~\footnote{\scriptsize{\url{https://bit.ly/stipjs} (accessed November 2017)}}, the output of the transformation is JavaScript code for both client and server side. 
We implemented our own communication library to perform remote procedure calls, together with a layer for replicated data on top of it. 
Listing~\ref{lst:stipchat} demonstrates the multi-tier code for a chat application. 
We make use of replicated data (line~\ref{lst:chatmessages}) and the built-in HTML templating system Redstone to create our reactive user interface.
Inside the HTML code we can access JavaScript variables and functions by wrapping JavaScript expressions in double curly braces.
Listing~\ref{lst:stipchatserver} and~\ref{lst:stipchatclient} illustrates how code for the data replication and reactive UI updates are inserted in the output of the tier-splitting process.

\begin{lstlisting}[basicstyle=\scriptsize\ttfamily, float,language=JavaScript,label=lst:stipchat,caption=A chat application in Stip.js,  escapeinside={\&}{\&}]
/* @server */
{
    /* @replicated */
    var messages = [];&\label{lst:chatmessages} &
}
/* @client */
{
    var name = "user" + (Math.floor(Math.random() * 9901) + 100);
    var msg  = "";
    function chat() {
        messages.push({name: name, message: msg});
        msg = ""; }
}
/* @ui */
{{#each messages}} &\label{lst:chatuimessages} &
    p {{name + " says: " + message}} &\label{lst:chatuivars}&
input[value={{name}}]
input[value={{msg}}][placeholder=Message]
button[@click=chat]#send Send
\end{lstlisting}

\begin{minipage}[t]{.46\linewidth}
\captionsetup{width=.8\textwidth}
\begin{lstlisting}[basicstyle=\scriptsize\ttfamily, language=JavaScript,label=lst:stipchatserver,  caption=Resulting server side code for Listing~\ref{lst:stipchat}, escapeinside={\&}{\&}]
var server = new ServerData(app, 3000);
var messages = server.makeReplicatedObject('messages', []);
server.expose({});
\end{lstlisting}
\end{minipage}%
\hfill%
\begin{minipage}[t]{.48\linewidth}
\captionsetup{width=.8\textwidth}
\begin{lstlisting}[ basicstyle=\scriptsize\ttfamily, language=JavaScript,label=lst:stipchatclient, caption=Resulting client side code for Listing~\ref{lst:stipchat}, escapeinside={\&}{\&}]
var client = new ClientData('http://localhost:3000');
var name = 'user' + (Math.floor(Math.random() * 9901) + 100);
var msg = '';
var messages = client.makeReplicatedObject('messages', []);
REDSTONE.updateVariable('name', name);
REDSTONE.updateVariable('msg', msg);
function chat() {
    messages.push({name: name, message: msg});
    REDSTONE.updateVariable('messages', messages); }
\end{lstlisting}
\end{minipage}

In the work presented in this section code blocks are coupled to tier-specific placement annotations (\textsc{@client} and \textsc{@server}). However, in this paper we advocate that placement should be decoupled from the source code into a separate configuration step. Our approach facilitates this decoupling because it relies on static analysis to determine at compile time what calls are between different tiers. Nothing in the source code distinguishes local from remote calls and only at compile time our transformation tool decides what calls should be transformed into their asynchronous counterparts. Furthermore, we show that this decoupling of placement annotations and tier-agnostic slices improves the flexibility of the resulting application. Different crosscutting concerns such as performance, offline availability, security, etc. that require different configurations can be easily accommodated. On top of that we present a recommender system that tries to calculate the optimal configuration and suggests refinements based on the results of that configuration.

\subsection{Untangling tier placement from web slices}

A number of cross-cutting concerns cannot be expressed by means of a simple annotation in the source code. 
For example, when a developer wants to improve the offline availability of his or her application, this can have an impact in various locations the code base. For example, a developer might choose to replicate a collection, but this will also have an impact on all the code that is using that collection. Meanwhile, the offline availability is improved but the overall security of the application might be lowered. There is no way to visualise or measure how well a certain configuration of the application handles these various concerns.
Generally speaking, handling cross-cutting concerns leads to code scattering and tangling between the tiers. 
Maintaining the application becomes harder as well: because of code duplication the code on the client and server tier handling this concern should both be maintained.

For this paper, instead of taking \textsc{@client} and \textsc{@server} slices as building units if the web application, we decouple the slices from their tier specification.
Slices are a unit of code that have a unique name and can be mapped to a tier in an external specification. 
Two new location-based annotations are thus added: \textsc{@slice} to define a new slice and \textsc{@config} to map slices to a certain tier.

\begin{lstlisting}[basicstyle=\scriptsize\ttfamily,language=JavaScript,label=lst:slices,  caption=Example of slices and configuration, escapeinside={\&}{\&}]
/* @config data : server, browser: client  &\label{lst:config} &
   @slice data */ &\label{lst:defineslice1} &
{
  /* @replicated */
  var tasks = [];
  /* @replicated */
  function Task(name, priority) {
  	this.name = name;
	this.priority = priority;
	this.status = -1;  }
}
/* @slice sorting */ &\label{lst:defineslice2} &
{
    function sortTasks () {}
}
/* @slice statistics */ &\label{lst:defineslice3} &
{
    function getTaskStats () {}
}
/* @slice browser */ &\label{lst:defineslice4} &
{
  function displayTasks () {
      sortTasks();
      tasks.forEach(function (task) { /* display the task */ })
  }
  function displayTasksStats() {
    var stats = getTasksStats();
    /* Generate and display graph */ }
}
\end{lstlisting}

Listing~\ref{lst:slices} shows how we could implement the part of the Uni-corn application that keeps track of the tasks of the student. 
We omitted the bodies of the functions for brevity.
We define four slices (on line~\ref{lst:defineslice1},~\ref{lst:defineslice2},~\ref{lst:defineslice3} and~\ref{lst:defineslice4}) and give them each a name: \emph{data}, \emph{sorting}, \emph{statistics} and \emph{browser}.
Each slice has its own block statement that consists of the code belonging to that slice.
In this example the \emph{data} slice is responsible of declaring the replicated data, the \emph{sorting} slice defines a function to sort the tasks based on each task's priority.
The \emph{statistics} slice has a function that calculates how many tasks are finished, in progress or have yet to be started.
The \emph{browser} slice has a function to display the tasks and the task statistics, after making sure the tasks are sorted.

\begin{figure}
\centering
 \includegraphics[scale=0.45]{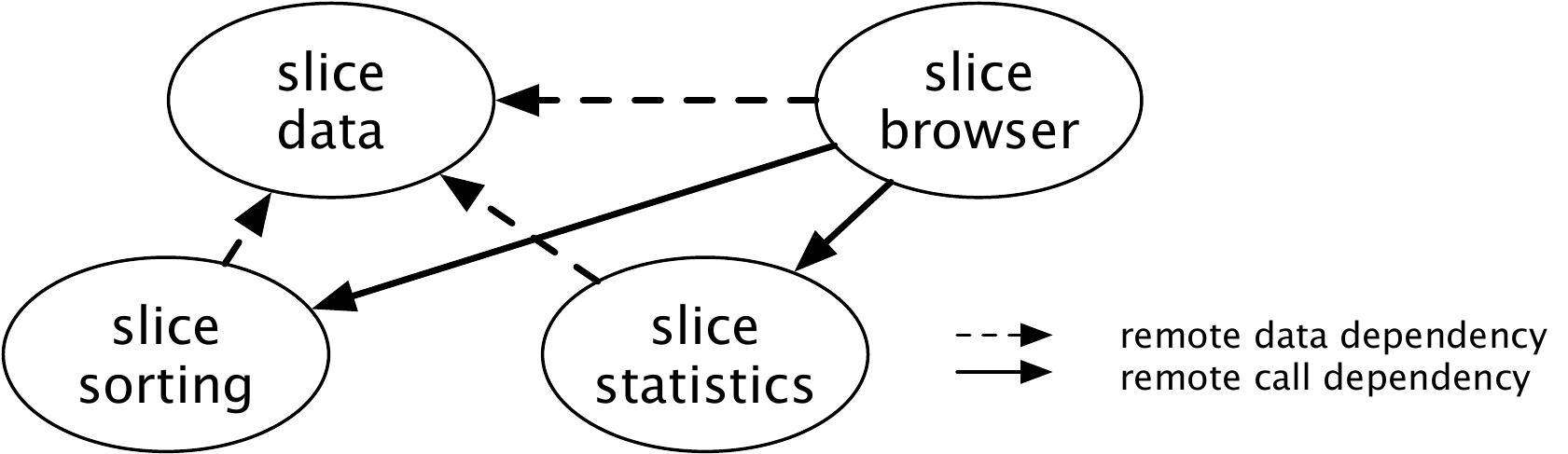}
\caption{Dependencies between the slices}
\label{fig:slicesgraph}
\end{figure}

Because our approach is based on a program analysis a dependence graph can be constructed that enables a visualisation of the dependencies between different slices, see Figure~\ref{fig:slicesgraph}.
It is actually a PDG as introduced in Section~\ref{sec:background} where the Distributed Component nodes (thus the slices) are collapsed.
Whenever a call or data dependency crosses the boundaries of a slice, it becomes a remote dependency regardless of the location of slices.
As can be seen on the graph, the \emph{browser} slice has only outgoing dependencies.
The \emph{data} slice on the other hand is a supportive slice: it has only incoming dependencies.

The configuration of the slices (Listing~\ref{lst:slices}, line~\ref{lst:config}) gives a fixed placement for the \emph{browser} and \emph{data} slice.
The placement of the \emph{browser} slice is fixed and must be on the client tier because it updates the user interface in the browser. The placement of the \emph{data} slice is also fixed and must be on the server tier because we want our server to manage a centralised set of tasks.
The slices that have no fixed placement in the configuration (in this case statistics and sorting) are not bound to a particular tier and their placement can vary depending on the output of our recommender system.

Placing the slices on a certain tier is the task of the recommender system. 
This component is responsible for taking a crosscutting concern into account when assigning slices to tiers.
For example, one system can try to maximise the offline availability of an application while another can try to optimise the memory usage.
The result of the recommender system is an allocation of every slice to a tier.
However, how well our recommender system can decide on a placement heavily depends on the amount of slices and their granularity. For this reason the recommender system has an additional step in which it reports back to the developer about potential refinements of the source code.
Given a placement for every slice, considerations for slice refinements are given to the developer.
By following these considerations, the developer can potentially improve how well a cross-cutting concern is handled within his application.
In the following sections we discuss the tier assignment and slice refinement process.

\subsection{Search-based tier assignment to user-defined slices}
\label{sec:search-based-tier-assignment}
The recommender system must be applied to allocate the unplaced slices to a tier, given one or more crosscutting concerns.
As an input it takes the configuration of fixed slices and the multi-tier program and as output it gives a mapping where every slice is assigned to a tier.
The system can resort to several techniques to calculate the most ideal placement: a dependence-driven analysis, code instrumentation, etc.

Because with each additional slice the amount of possible configurations increase exponentially it is impossible to consider every potential configuration. Therefore we opted for a genetic search algorithm in order to reach, or at least approximate, the optimal placement to satisfy a crosscutting concern.
Genetic search algorithms have their origin in natural selection: starting from a population of individuals, called a \emph{generation}, from which the fittest candidates are selected to form a new generation by performing mutations and combinations (crossovers). 
This process repeats until a certain end condition is satisfied or a certain number of generations have been computed. 

The only required input for a genetic search algorithm is a starting population, a fitness function and method for generating offspring (mutations) from an existing population. The only part of that input that is specific to a certain cross-cutting concern is the fitness function. The following section gives a brief overview of each component of the search algorithm and gives an example fitness function for optimising offline availability of an application. However, this fitness function can be altered to focus on other concern(s).
Currently, the search we apply is single-objective: this means that during the search we optimise only one objective. 
We leave the case where multiple objectives are optimised at the same time and trade-offs between possible conflicting objectives as future work.

\paragraph{Generation}
The individuals of the population are a mapping of unplaced slices to a tier, in this case client, server or both.
Every iteration of the genetic search produces a new generation of a fixed size, originating from the previous generation.
The initial seed for the search algorithm is a mapping where every unplaced slice is assigned to a random tier.
Because every unplaced slice has three options, we have a search space of order $3^n$, with n the number of unplaced slices.
In the selection process we discuss how we reduce this search space.

\paragraph{Fitness function}
Because a single-objective search is conducted, we have a single fitness value that the algorithm tries to maximise.
For this example, our aim is to get an high offline availability, thus our fitness function calculates a real number in the [0,1] range that represents how offline available the current configuration is. Before we define the fitness function, we need to able to define \emph{how offline} a certain slice, placed on a given tier, is.
\small
\[\text{OFFLINE\,\%}(slice) = \frac{localCalls(slice)} {calls(slice)} \]
\normalsize
For this we calculate the number of local calls performed in the slice, which are calls to functions defined in the slice itself or to slices placed on the same tier.
This is divided by the total number of calls performed in that slice, this number thus includes calls to functions defined in slices placed on other tiers.
We do not claim that this is an ideal function for measuring offline availability of a certain slice. However, our experimental results have shown that this function gives a good enough approximation.

Our fitness function regards every slice and calculates a weighted mean of every $OFFLINE\,\%$ score for every slice. 
This way slices that perform a large amount of calls have more influence on the final offline availability score of the configuration.
\small
\[\text{OFFLINE\,\%}(program) = \frac{\displaystyle\sum_{i=1}^{\substack{\# slices\\ program}} OFFLINE\,\%(slice_i) * calls(slice_i)} 
							{\displaystyle\sum_{i=1}^{\substack{\# slices\\ program}} calls(slice_i) } \]
\normalsize

Note that the calculation of this fitness function is partly based on results from our static analysis tool. Whether a call is local or remote is part of the PDG generated by the analysis. Our current implementation uses the JIPDA abstract interpretation~\cite{nicolay2015detecting} framework. Designing such a fitness function thus requires expert knowledge of our framework and we envision that a number of these are developed upfront for each cross-cutting concern that can be considered.

Note that, both the chosen fitness function as well as the strength of the analysis have an impact on the result of executing the algorithm. For example, the current tool we employ for static analysis of the JavaScript code does not fully take into account remote calls from within a program loop. When the client performs a remote call to the server in a loop, this call will be represented in our PDG as only one remote call dependency. A recommender system that tries to minimise inter-tier communication would currently be unable to differentiate between a remote call inside and outside of such a loop.

\paragraph{Selection Process} 
The creation of a new generation is achieved by performing selections and mutations on the current generation.
A mutation consists of randomly picking a slice from the placement mapping and assigning it to a random tier.
A selection undertakes a tournament selection, which randomly picks a number of mappings from the current generation and returns the one wth the best fitness value.

However, not every generated mapping is a valid one. 
Consider, for example, a slice that defines a function that is called by a (fixed) server slice and a (fixed) client slice. 
The slice itself performs no calls to functions defined in another slice. 
If we place this slice on the client tier, we get a higher offline availability, because the client tier now \emph{gains} local calls instead of making remote calls.
However, the server slice that called the function as well now has to perform a remote call to a slice on the client slice.
At runtime, this means that the server instance must pick one or more connected clients to call this function and wait for the result. 
Because it is not clear whether the local call should be a broadcast to all clients or to a specific client, this is not a valid placement of the slices.
If the client performs the call on all clients, what should happen with possible return values from the clients, or in case of a specific client, which client is selected?
Please note that the \textsc{broadcast} and \textsc{reply} annotations from Table~\ref{tbl:stip:annotations} can be used for server-to-client calls. 
If the call in the server slice is annotated with one of these annotations the program would be valid.

These invalid placements could have a higher fitness score than valid placements, and this is why the selection process is responsible of not selecting these configurations as a base for the next generation.

\paragraph{End condition}
We run the genetic search until either a mapping has been found that produces a fitness equal to 1 or when a certain threshold of generations has been reached.
In the latter case, the fittest individual of the latest generation is returned.
Because a genetic search is non-deterministic, the result of different runs of the algorithm can yield different placements of the slices. The more generations are considered, the better the final placement will address the concern.

\subsection{Recommending slice refinements after assignment}
After a placement configuration is calculated by the genetic search algorithm an additional step is taken.
The result of this step is advice that the programmer can consider to integrate such that the resulting web application can score better on the concern that is under consideration.

Our recommendation system currently only supports suggestions for improving offline availability. However, in the future other recommendation functions can be considered. 

For offline availability, in the recommendation phase, we consider all slices and every data or function declaration made within those slices.
Data declarations can either be tied to a tier or be replicated. 
Generally speaking, replicating data has a positive effect on the offline availability of a web application as it makes sure that the client has a local copy of the data.
Therefore, we look for data declarations that are not indicated to be replicated (by means of an annotation), but that are often used in functions that are called remotely by another tier. 
For example, we declare data on the server side and define a number of accessors for retrieving and manipulating that data.
If these functions are heavily called by the client tier, it might be a better solution to replicate that data such that the client has to communicate less with the server.

To detect this, we use the dependencies present in the PDG. 
We look at every declaration and follow its data dependencies to statements in the same tier. 
If these statements reside in a function that has more remote calls than local calls, we give the programmer the advice to replicate the data.

Function declarations might be a good candidate to move to another tier, or to a new slice, if it is called \emph{more} by another tier than the one it is defined in.
We define \emph{more} here as the percentage difference between the number of local calls and the number of remote calls to that function is bigger than a certain threshold. To calculate the number of (remote) calls we also use the dependencies in the PDG.
If we observe functions that fulfil this requirement, we advice the programmer to move it to a new slice. 
This way, the recommender system can figure out on which tier(s) the function definition should end up.

Listing~\ref{lst:advicereport} shows an advice report for the code given in Listing~\ref{lst:stipadvice}.
The code is (part) of an implementation of the Uni-corn application where we define two fixed slices: data and browser placed on server and client tier respectively.
The server slice declares data collections, as JavaScript arrays, and functions to retrieve and manipulate these collections.
The client slice is responsible for the user interface: it defines a set of functions that are called from the HTML part of the application.
The client code uses these server-defined functions heavily to retrieve and update the collection's according to the user's input.
As a consequence, a lot of remote communication transpires and when no connection is within reach, the application becomes impractical and useless for the user.
A solution would be to replicate the data, such that the client has a local copy available as well.
The helper functions need to be available on the client side as well, so they should be moved to an (unplaced) new slice.
The recommender system makes sure that the new slice is put on client and server side, thus reducing the remote communication.
This scenario is what the report suggests as well, as can be seen from the output of the recommender system (Listing~\ref{lst:advicereport}). 

\begin{minipage}[t]{.48\linewidth}
\captionsetup{width=.98\textwidth}
\begin{lstlisting}[basicstyle=\scriptsize\ttfamily, language=JavaScript,label=lst:stipadvice,  caption=Two fixed slices, escapeinside={\&}{\&}]
/* @config data:server, browser: client 
   @slice data  */
{
    var meetings = [];
    var tasks = [];      
    function getMeetings() {}
    function getTasks() {}
    function addMeeting(name, date) {}
    function addTask(name, priority) {}
}
/* @slice browser */
{ /* Display tasks, meetings and event listeners  */ }
\end{lstlisting}
\end{minipage}%
\hfill%
\begin{minipage}[t]{.47\linewidth}
\captionsetup{width=.9\textwidth}
\begin{lstlisting}[basicstyle=\scriptsize\ttfamily, numbers=none, language=bash,label=lst:advicereport,  caption=Advice report for Listing~\ref{lst:stipadvice}, escapeinside={\&}{\&}]
$ node main.js examples/uni-corn.js
  Application level of offline availability: 10\,%
  Consider making following declarations replicated
        - var meetings
        - var tasks
  Consider moving following functions to new slice:
        - getMeetings
        - getTasks
        - addMeeting
        - addTask
\end{lstlisting}
\end{minipage}

In certain cases it would be beneficial to detect these declarations and move them automatically to a new slice.
However, if these declarations are defined in a tier with a fixed placement, it might be done on purpose by the programmer.
If the recommender system automatically makes declarations replicated by adding an annotation on top of it, we might break some validation or security checks that are performed on the functions that manipulate the data. 
Because the client can never be trusted, the programmer's intention could be that the data must remain on the server side and is only accessible through these functions.
For this reason we only give advice and don't perform transformations on the fixed slices.

\section{Evaluation}
\label{sec:evaluation}
Our evaluation aims to show that a genetic search algorithm can recommend a good placement strategy for slice-based web development and that our recommender system can further improve upon this by recommending refinements to existing slices. We currently only support offline availability as a cross-cutting concern, but our approach is applicable for other concerns as well.

In order to validate our approach we seek to answer the following research questions:
\begin{itemize}[leftmargin=1cm]
\item [RQ1] Can the recommender system identify a location assignment of the slices and recommendations that lead to a web application with a high(er) offline availability?
\item [RQ2] Does the slice granularity have an impact on the effectiveness of the recommender system?
\item [RQ3] How does our approach compare to other multi-tier approaches with respect to the offline availability of the resulting web applications?
\end{itemize}

We answer these questions based on an evaluation of several implementations of the Uni-corn application as presented in Section~\ref{sec:motivatingexample}.
The application represents several characteristics of a modern, rich internet application: it is a single page app where updates on the data are automatically reflected in the user interface as well, without the need to refresh the page. 
Moreover, the user is able to update, add and remove tasks or meetings in an offline setting as well.
On top of this, the application uses a number of JavaScript libraries.

Because offline availability is one of the defining characteristics of today's web applications, we use this concern as our main goal for the recommender system.
This means that in each of the following evaluations we measure and try to maximise the offline availability of the Uni-corn application.

The code used for the evaluation is publicly available\footnote{\scriptsize{\url{https://github.com/lphilips/multitier-approaches} (accessed November 2017)}}.

\subsection{Recommender system for offline availability}
\begin{figure}
\centering
\includegraphics[scale=0.35]{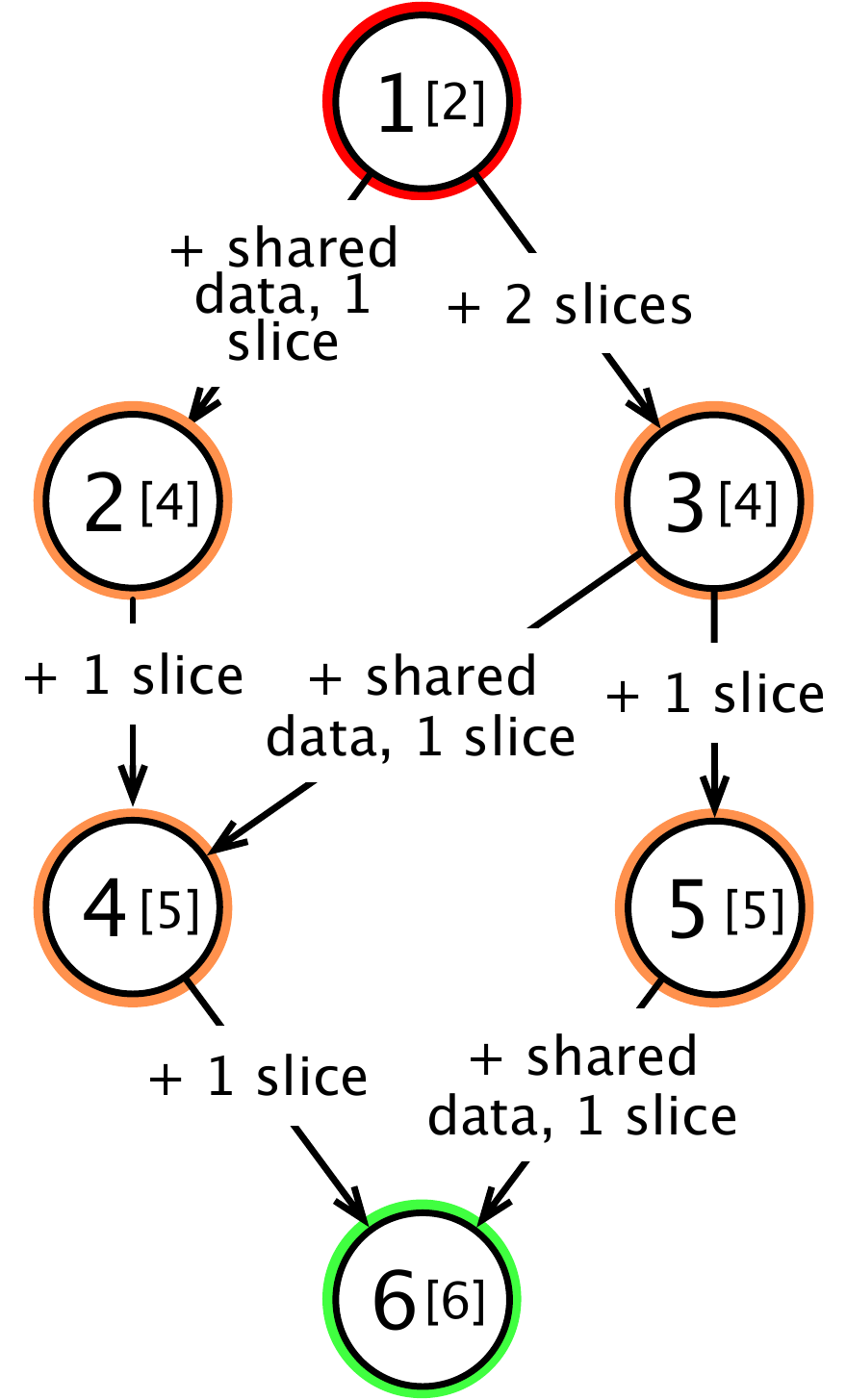}
\caption{Relations between the different versions of Uni-corn app}
\label{fig:eval-versions}
\end{figure}

To evaluate the strength of our recommender system for enabling offline availability we have implemented several versions of the Uni-corn application.
We start from a first version that contains only two slices with a fixed placement: one on the client tier, the other on the server tier.
In each step we follow all of the recommendation for refinements suggested by our recommendation system. This means we either follow the advice to add data sharing annotations to certain declarations or move certain functions to new slices.
We keep doing this until we end up with a version that receives a fitness value of 100\,\% (In our example, a fitness of 100\,\% is achievable with the correct placement strategy. However, this is not necessarily true for all applications).
In total we have six versions, Figure~\ref{fig:eval-versions} shows how these versions evolve and Table~\ref{tbl:eval-unicorns} lists several properties of the code, such as the number of annotations categorised according to Table~\ref{tbl:stip:annotations}.

Every version of the application has at least two slices that are tied to a certain tier and thus have a fixed placement. In the final version these slices contain the smallest subset of the source code that is bound to either the server or client tier.
The recommender system is in charge of coming up with a placement of the remaining unplaced slices.

\begin{table}
\centering
\caption{Versions of Uni-corn app}
\scriptsize{
\begin{tabular}{c c c c c c}
\hline
ver. & \# slices & \# fixed slices &  @placement  & @communication & @sharing\\ \hline 
1 & 2 & 2 & 3 & 5 & 0	\\
2 & 4 & 2 & 5 & 1 & 6	\\
3 & 4 & 2 & 5 & 5 & 0	\\
4 & 5 & 2 & 6 & 1 & 6	\\
5 & 5 & 2 & 6 & 5 & 0	\\
6 & 6 & 2 & 7 & 0 & 6	\\  \hline
\end{tabular}}
\label{tbl:eval-unicorns}
\end{table}%

For this evaluation we run the genetic search 100 times for each version and analyse the outcomes. 
Table~\ref{tbl:eval-placement} summarises for each version the minimum, maximum and median number of slices placed on every tier (denoted as minC, maxC and medC respectively for the client tier; minS, maxS and medS for the server tier; minB, maxB and medB for both tiers). 
It also lists the fitness value for every version, which is the same for the 100 runs, because the search always selects the highest number, but the configuration of the slices can differ between runs.
We also specify the number of data declarations that could be replicated (data adv.) according to the recommender system and the number of functions that could be moved to a new slice (slice adv.).

The genetic search has the following configuration: the probability for crossover and mutation are both equal to 0.6 ; the maximum number of generations is 300; the size of each generation is 30 individuals. 
The fitness function and other characteristics of the genetic search are the same as described in Section~\ref{sec:search-based-tier-assignment}.
We use a genetic search library for JavaScript\footnote{\scriptsize{\url{https://subprotocol.com/system/genetic-js.html} (accessed November 2017)}}.

\begin{table}
\caption{Results of 100 runs of the recommender system}
\centering
\newcolumntype{s}{>{\hsize=.4\hsize}X}
\scriptsize{
\begin{tabularx}{0.98\textwidth}{s s s s s s s s s s s X Y Y}
\hline
ver. & gen. & medC & minC & maxC & medS & minS & maxS & medB & minB & maxB & offline\,\% & data adv & slice adv\\ \hline 
1 & 300 & 1 & 1 & 1 & 1 & 1 & 1 & 0 & 0 & 0 & 10,52\,\%	& 6 & 13\\
2 & 300 & 2 & 1 & 2 & 1 & 1 & 1 & 1 & 1 & 2 &  66,67\,\%	& 0 & 6 \\
3 & 300 & 1 & 1 & 2 & 1 & 1 & 1 & 2 & 1 & 2 & 68,42\,\%	& 5 & 5\\
4 & 300 & 2 & 1 & 2 & 1 & 1 & 1 & 2 & 2 & 3 & 85,71\,\%	& 0 & 3\\
5 & 300 & 2 & 1 & 3 & 1 & 1 & 1 & 2 & 1 & 3 & 68,42\,\%	& 5 & 5\\
6 & 1 & 2 & 1 & 3 & 1 & 1 & 1 & 3 & 2 & 4 	& 100\,\% 		& 0 & 0 \\  \hline
\end{tabularx}}
\label{tbl:eval-placement}
\end{table}%

From Table~\ref{tbl:eval-placement} we can see that following up on the advice has a positive impact on the offline fitness score of the application.
Version 6 has a fitness value of 100\,\%, because more slices have been added and data has been replicated.
The versions that have a higher fitness score also place more slices on both client and server tier.
Table~\ref{tbl:eval-unicorns} indicates that adding annotations for replicated data reduces the number of communication annotations that must be used.

From this evaluation we can answer the first research question (RQ1).
We have successfully used the recommender system to evolve the initial web application to one where the offline availability according to the recommender system's report is higher than that of the initial application.
Integrating the recommender system's advice leads to an increase of the fitness score.
The distribution plan computed by the recommender system that focuses on offline availability clearly prefers to put slices on the client tier or duplicates them between client and server. 
The different runs of the genetic search on the same version lead to distribution plans that might differ in the number of slices that are put on the client tier or on both.

\subsection{Slice granularity}
Because the recommender system decides on a distribution plan based on the unplaced slices, its effectiveness depends on the programmer to make a program with enough slices. 
In the ideal case, the fixed slices only contain code that is restricted to one of the tiers and all remaining code is put inside unplaced slices.
Table~\ref{tbl:eval-placement} shows that the version of the Uni-corn application with only two slices has a low offline availability score and receives a considerable amount of feedback.

For this reason we implemented an extension of our tier splitter that automatically follows up on the advice.
For example, if the recommender system indicates that a declaration should be shared, we adapt the original AST of the program such that it now has an \textsc{@replicated} annotation.
If the recommendation indicates that a function should be moved to a new slice, we remove the function from its original slice, add a new slice to the program with no fixed tier and add the function as only expression in that slice. 

We ran the same experiments as before for the six versions of the Uni-corn app, but in between runs we automatically incorporate the suggestions made by the recommender system, the results are given in Table~\ref{tbl:eval-placementauto}.

\begin{table}
\caption{Results of automatically integrating the advice}
\centering
\scriptsize{
\begin{tabular}{c c c c}
\hline
ver. & max. for OFFLINE\,\% found after & original nr. of slices &  nr. slices in end configuration  \\ \hline 
1 & 1 run & 2 & 15	\\ % 13 + 2
2 & 1 run & 4 & 10	\\ % 6 + 4
3 & 1 run & 4 & 9	\\ % 5  + 4
4 & 1 run & 5 & 8	\\ % 3 + 5
5 &  1 run & 5 & 10	\\ % 5 + 5
6 &  0 runs & 6 & 6	\\   \hline % 0 + 6
\end{tabular}}
\label{tbl:eval-placementauto}
\end{table}%

The results make clear that integrating the advice automatically produces the desired raise in offline availability immediately.
In the first version, where only two slices were initially present, 13 new slices were introduced, each defining a function that was advised to be moved to a new slice (as can be seen in table~\ref{tbl:eval-placementauto}). 
Following the advice concerning replicated data does not result in a new slice, but merely adapts the original code.

We can conclude from this table that automatically integrating the advice into the code results in a higher offline availability.
This answers the second research question: slice granularity has an impact on the effectiveness of the recommender system.
More slices leaves more room for the recommender system to optimise the distribution plan, while few and big slices may lead to inferior results.

However, the automatic integration adds a lot of new slices or annotations to the code, thus making it harder for the developer to map the transformed code back to the multi-tier input.
As mentioned before, we do not perform these changes automatically but rather give them to the developer as feedback.
The reason is that code that was originally in a fixed slice should not always be moved automatically, because the moved portions could contain code that cannot be shared with or moved to another tier.

\subsection{Evaluation comparison of the three multi-tier approaches}
In order to compare the three approaches to multi-tier programming we implemented the Uni-corn application using specific frameworks representative for each approach. 
For transformation-based multi-tier programming we have our own implementation in Stip.js (we use the final version in Table~\ref{tbl:eval-unicorns}). 
For library-based approach and language-based approach we implemented the Uni-corn application in the Meteor framework and the Hop.js language respectively. 
We chose these two platforms specifically, because they both use JavaScript as a host language and are both actively maintained. 
Because all three platforms have support for JavaScript and all the utilised libraries, we can reuse the program logic for every implementation. 
For that reason we can implement the Uni-corn application with three different technologies, without drastically having to change the used libraries or program logic.
We implemented the three versions with the libraries or language constructs that are available, without writing our own extensions or packages.
The code of the three versions is compared by means of number of source lines of code (SLOC) and the number of tier-switching or tier-specific annotations or calls to libraries.
This gives us an insight how much effort is required from the programmer to enable remote communication, data sharing or getting offline availability.
The runtime characteristics are used to measure the actual offline availability of the resulting applications by looking at the actual remote communication performed between server and client, the amount of data transferred, etc.

\paragraph{Code characteristics}
For every implementation we count the total SLOC for server, client and UI code as well. 
For the Stip.js version we have code that belongs to slices that are not assigned to a tier, so we count that code as \emph{undecided}.
As can be seen, the three versions fall into the same category for the total number of SLOC.
However, the Hop.js variation focuses more on the server part, while the two other implementations have a small server setup.
Please note that for the Stip.js version this is not the only code for the server tier: parts of the undecided code could be server-side code as well.
We also enumerated the number of annotations: these are the tier-switching operators \texttt{\~}and \texttt{\$} for Hop.js and any annotation from Table~\ref{tbl:stip:annotations} for Stip.js.
For the library-based version there are no annotations present, because all the tier information is hidden away in the usage of the framework's libraries.
For that reason we counted the number of calls to libraries for data sharing, reactive updates, and so on.
The Hop.js version has significantly more annotations to escape from the server to client tier and vice versa.

\begin{table}
\caption{Code characteristics of the three multi-tier Uni-corn applications}
\centering
\scriptsize{
\begin{tabularx}{0.95\textwidth}{X  X  X  X  X  Y  Y}
\hline
& SLOC & SLOC client & SLOC server & SLOC ui & SLOC undecided & annotations/ library calls \\ \hline
Meteor & 590 & 300 & 27 & 263 & 0 & 19 \\
Hop.js & 579 & 239 & 360 & & 0 & 28 \\ 
Stip.js & 468 & 115 & 30 & 192 & 131 & 13 \\ \hline
\end{tabularx}}
\label{tbl:eval-code}
\end{table}%

\paragraph{Runtime characteristics}
We evaluate the three multi-tier variants of the Uni-corn app through several scenarios. 
Because the app offers four services to the user, we first evaluate these services individually in an online and offline setting.
This means that we view, add and update meetings and tasks and view the schedule and charts. 
We measure the number of requests that are made, the number of failed requests and kilobytes transferred.
Then we look how many steps of the scenario succeeded: i.e., the application produces the correct results. 
For example, adding a new meeting should first of all not produce an error and the entered data should not be lost, even in an offline setting.
Moreover, the meeting should be added to the list of meetings and the user should be able to retrieve the details of the meeting to alter it later on.
The results of evaluating the different services this way can be found in Appendix~\ref{app:scenarioeval}.
We did the same for three scenarios, given in Table~\ref{tbl:overviewscenarios} in Appendix~\ref{app:scenarioeval}, that combine more of the services, e.g., adding a task and viewing the progress charts later on.

To measure these characteristics we used Chrome Developer Tools on Google Chrome browser, version 56. 
To inspect network traffic for the Meteor version we used Meteor Devtools\footnote{\scriptsize{\url{https://github.com/bakery/meteor-devtools} (accessed November 2017)}}, because Meteor uses its own communication protocol that can't be inspected via Chrome Developer Tools.

\begin{table}
\caption{Results for the three scenarios from Table~\ref{tbl:overviewscenarios}}
\centering
\scriptsize{
\begin{tabularx}{0.85\textwidth}{c *{7}{Y}}
\hline
 & number of requests & failed requests & $\frac{\text{failed requests}}{\text{total requests}}$ & kb transferred & number of steps & $\frac{\text{successful steps}}{\text{total steps}}$  \\ \hline
\multicolumn{7}{c}{Scenario 1 } \\
Meteor & 6  & 2 & 33,3\,$\%$ & 4,63 & 7 & 100\,\% \\
Hop.js & 10 & 4 & 40\,$\%$ & 46,2 & 7 & 71,4\,\%\\
Stip.js & 16 & 0 & 0\,$\%$ & 2,959 & 7 & 100\,\%\\ \hline

\multicolumn{7}{c}{Scenario 2} \\
Meteor & 6 & 2 & 33,3\,$\%$ & 4,83 & 7 & 100\,\%\\
Hop.js & ? & ? & ?\,$\%$ & ? & 7 & ?\,\% \\
Stip.js & 24 & 0 & 0\,$\%$ & 3,769 & 7 & 100\,\%\\  \hline

\multicolumn{7}{c}{Scenario 3} \\
Meteor & 6 & 0 & 0\,$\%$ & 1,185 & 6 & 100\,\% \\
Hop.js & 10 & 7 & 70\,$\%$ & 10 & 6 & 0\,\%\\
Stip.js & 8 & 0 & 0\,$\%$ & 1,043 & 6 & 100\,\%\\ \hline
\end{tabularx}}
\label{tbl:eval-scenarios}
\end{table}%

As can be seen from Table~\ref{tbl:eval-scenarios}, both Meteor and Stip.js perform the different steps correctly, even in an offline setting. 
This means that no errors were shown to the user, no data loss occurred, and the user was actually unaware of the lost connection.
The reason is that Meteor has replicated Collections that buffer every change on the client side to send to the server when the connection is restored.

A significant change between the different approaches can be seen in the number of requests performed: the Stip.js version performs more requests in the first two scenarios.
This is because the run-time library for remote communication and data sharing\footnote{\scriptsize{\url{https://github.com/lphilips/asyncCall-with-shared-data} (accessed November 2017)}} works on a fine-grained level when changes to a replicated are made: every property change on a replicated object is sent to the server. 
For example, two changes are communicated when adding an element to a replicated array: one for the index on which the object is added and one for the length that has changed.
The library we use is smart enough to keep track of the connection with the server through a heart-beat system, such that communication is automatically buffered when no connection is present and thus produces no failed requests.
Another consequence of having replicated data available on the client is that fewer kilobytes need to be transferred.

One scenario in the Hop.js version could not be performed (indicated by a question mark), because the Hop.js compiler cannot integrate the JavaScript library that renders the calendar.
We notified the authors of this issue, but the problem persists at the time of writing. 

In the Hop.js version the HTML content is rendered on the server side, this resulting in larger data size transferred.
At the time of writing, Hop.js has no support for data replication or reactive UI updates, so updates and retrieving the data in an offline setting do not work.
Both the Meteor and and Stip.js versions only have to communicate the changes made to the replicated objects and the user interface is updated automatically.

This table shows that certain multi-tier approaches excel at optimising certain crosscutting concerns.
However, focusing on other crosscutting concerns often force the developer to extend the framework or language by hand.
The Meteor framework scores highly on offline availability for the reason that its built-in Collections are replicated to every client.
The Hop.js language has no support yet for data replication and offline availability, thus leading to more communication to the server that fails when no connection is at one's disposal.

%RQ3 How does our approach compare to other multi-tier approaches with respect to the offline availability of the resulting web applications?
Both the runtime and code characteristics of the three versions of the Uni-corn application gives us an answer to the third research question.
For this application the code characteristics are comparable, although Stip.js requires less lines of code and annotations.
On the other hand, Stip.js is not tailored towards runtime optimisation, resulting in more actual web requests that are performed.
On the other hand, because the resulting application focuses on offline availability, less data is transferred with each request.
The language-based multi-tier approach is not tailored towards offline availability, resulting in all data in the shared collections to be transferred each time.

\paragraph{Threats to validity}
For the motivating example, this evaluation shows that the recommender system can successfully help the developer to achieve full offline availability. However, we were also able to manually achieve full offline availability for the motivating example. Whether or not our recommender system is better at finding placement strategies than a manual approach, or whether our recommender system can help developers find an optimal strategy faster would require an extensive user study for much larger applications. Additionally, the evaluation currently only covers offline availability as a single-objective optimisation. To show that the approach is also applicable to other single- and multi-objective optimisations would require additional exploration. Finally, the evaluation compares the approach with two other tierless programming platforms. A complete comparison that covers a wider range of platforms would strengthen our comparison with the related work.

\section{Related Work}
\label{sec:relatedwork}

Current techniques that infer a distributed placement mostly focus on minimising the communication cost between the distributed tiers of the application.
When implemented on top of a new language, the placement analysis is often guided by primitive operations: e.g., \emph{all \texttt{gui\_...} operations must be allocated on the client tier}. 
From there on the location analysis propagates throughout the program and assigns each operation/declaration to a set of possible tier locations.
We give an overview of approaches that can be used for web applications, but most of them handle distributed applications in general.

\subsection{Multi-tier Approaches}
We classify multi-tier approaches in three main categories: language-based, library-based and transformation-based approaches.
Library-based and transformation-based approaches target a general-purpose language instead, thus facilitating the reuse of existing developer tools.
We now discuss multi-tier approaches that support a location analysis to assign expressions to a tier.

The multi-tier language Opa~\cite{opa:2013} supports annotations to indicate where functions or data should be located. 
In addition, Opa's slicer takes the type information and makes a call graph to decide which function/values end up on what tier. 
By default values appear on both tiers whenever possible, but are constrained by security concerns or the behaviour they execute. 
DOM-related operations end up on the client whereas database operations are always placed on the server tier. 
It is not clear how the slicer makes the decision and what the goal of the placement algorithm is.
In contrast with Opa, our approach is extensible such that the programmer can try out different location strategies.
These location decisions are deeply embedded in Opa's slicer and the programmer must first test the slicer's result before adding more specific annotations to influence the decisions made by the slicer.

Distributed Orc~\cite{distributedorc:2016} is a distributed extension of the Orc programming language.
The language introduces location transparency by not abstracting away distributed concerns, but by explicitly making local and remote semantics uniform. 
As a consequence, asynchronous calls and failure handling are consistent throughout the whole program, even for local operations.
Its location analysis uses the locations of data and decides when data is used whether it is by using a copy of that data, migrating the execution to another location or another manner. 
It is ongoing research at this point and the authors state that it remains an open question whether there is a profit in communication costs when using these optimisers. 
At the moment the programmer cannot give an initial or partial distribution specification and it is not clear which analysis would drive the distribution decisions. 

The placement inferencer for a client-server calculus presented in~\cite{Neubauer:2005, placementinference:2008} is guided by the fixed placement of primitive operators.
The location analysis allocates every operation to a set of locations and propagates these assignments through the program.
This technique does not require the programmer to give an initial seed to the placement decision algorithm, as many others, amongst which Stip.js, do require.
However, this approach is not applicable to general-purpose languages. 
For example, primitive operators to update the DOM could easily be renamed or this could be achieved by means of an external library.
It is a difficult task to automatically infer this and an initial seed is thus necessary when targeting general-purpose languages.

\subsection{Location Transparency}
Location transparency abstracts away the place of execution of certain parts of a distributed program.
Several languages and frameworks therefore hide the semantics of remote communication between these parts and support a uniform way of communication.
Location transparency has been criticised~\cite{Waldo94anote}, mainly because concerns such as latency, memory access, partial failures, concurrency, etc. should not be abstracted away from the programmer.
The reason is that remote calls have a fundamentally different semantics to local calls, and thus fundamentally different failure handling code must be provided by the programmer as well.

Many distributed languages support mobility of distributed components, thus decoupling the components in space by communicating in an asynchronous way.

Actors are an example of such decoupled components~\cite{kim1995efficient}.
Erlang~\cite{Armstrong:2007:PES} is an example of a programming language that supports location transparent actors.
Every actor has a unique address and can be accessed through an actor reference.
There is no differentiation between local or remote actor references, and all inter-actor communication is done via asynchronous message passing.

Modular design is a field of software engineering that is present in many programming languages and architectures.
Remote-OSGi (or R-OSGi)~\cite{Rellermeyer2007} uses centralised module management as a starting point for distributed Java applications.
Modules are the units of distribution that implement a service-oriented architecture.
Local and remote service invocation is indistinguishable, and communication failures are handled as local module events.

Our approach offers location transparency, because the execution location of parts of the program is decoupled from the code itself.
Just as other approaches we support a uniform way of communication, opting for local communication.
It is the burden of the transformation tool to transform local to remote communication.
However, issues such as failure handling are not completely abstracted away from the programmer, but are added by means of annotations.
These annotations work for communication that stays local and the communication that is transformed to remote calls.

\subsection{Automatic Partitioning}
Code partitioning is a process that computes a mapping from code partitions to the nodes of a distributed system.
This process can take several factors into account, such as the hardware characteristics of certain nodes, profiling information, etc.

Coing~\cite{Hunt:1999:CAD} is an automatic partitioning system for distributed applications that consist out of components.
The partitioning depends on a graph representation of the remote communication based on scenario-based profiling.
The profiling step gathers information while the user runs the application but an automated testing tool can be used as well.
All communication that crosses the boundaries of the components is instrumented, together with the amount of transferred data.
A graph slicing algorithm is then applied to split the application and come up with the most optimal placement of components. 
While the programmer can tweak the results, overriding the distribution decisions seems not possible.
Coign resembles the approach taken by Stip.js, but we use a static analysis instead of scenario-based profiling.
The recommender system of Stip.js could use the profiling and instrumentation step of Coign, possibly leading to a more accurate report.

Secure program partitioning~\cite{Zdancewic:2002:SPP} is a splitting process that protects confidentiality of data in a distributed application.
The splitter takes an annotated program and a set of trust declarations and produces programs that satisfy all security policies.
This automated partitioning enables the programmer to write a program independent of its distributed setting, but with strong guarantees about the flow information.
The partitioning has been extended with data and code replication~\cite{Zheng:2003:URP}, to increase the flexibility of the splitter but keeping the same guarantees meanwhile.
The emphasis of the work is on security and respecting user-defined security contracts. 
In contrast to our approach, expressing contracts for other crosscutting concerns is not supported.

\section{Discussion and Future Work}

Our approach employs a non-deterministic search algorithm in order to prune the search space for potential placement configurations. However, for applications with a small number of slices a deterministic approach could be used. In our example, the non-deterministic strategy was able to successfully get to an ideal placement with a fitness of 1. However, additional experiments are required to further strengthen our claim that a genetic search algorithm is a good fit for finding placement strategies. Also, additional experiments need to be conducted to validate whether this approach does not land in local optima when exploring the search space. We also currently have not experimented with different versions of the fitness function. 

We currently have only explored offline availability as a cross-cutting concern. However, we believe that our framework can easily be extended to other cross-cutting concerns such as memory usage, security, communication cost, etc. By adding the appropriate fitness function and extending the recommender in order to incorporate advice that is specific to that cross-cutting concern.

Our fitness function for offline availability now employs the results of our static analysis in order to evaluate different configurations. However, for other fitness functions, this can be changed to also include a runtime analysis or other information into account.

\section{Conclusion}
\label{sec:conclusion}
In most tierless programming languages or frameworks today communication between different tiers is marked with an explicit boundary between source code fragments. This means that developers are required to think upfront about the placement of the different components of their application. Moreover, several non-functional, cross-cutting concerns such as offline availability and security are impacted by this placement strategy. This means that changing the placement of certain code fragments in order to improve on some of these concerns over time becomes impractical. All of these concerns have to be taken into account in the initial design of the application. This leads to applications that are hard to maintain.

The main contribution of this paper is slice-based web development as a tier-agnostic development strategy. The placement of different slices onto the different tiers can be done separately. We show that such a configuration can be used to tackle several cross-cutting concerns without the need to significantly change the original source code.  Moreover, we show that a recommender system can be built to automatically suggest a placement strategy to optimise for certain concerns. Our recommender suggest a specific placement of the different slices onto the different tiers and gives advice for slice refinements. Such advice could be to add annotations in the code, move certain parts of the code to a new slice, etc.

Additionally, we provide a preliminary evaluation of our approach by applying it to our motivating example (the uni-corn app) for one specific cross-cutting concern, namely offline availability. We start from an implementation with the minimal number of slices, one for the client and one for the server and gradually incorporate the feedback from our recommender system. The results of our evaluation show that our recommender system can successfully help developers optimise for offline availability. We also assessed the slice granularity and what impact this has on the results.

To evaluate our approach compared to the state of the art in tierless programming, the uni-corn application is also implemented in a library-based and language-based approach. We scrutinised the runtime and code characteristics of the three implementations and how they handle the offline availability concern.
We conclude that the slices in combination with a recommender system allow the programmer to focus on the application logic and takes the burden away of distributing the code by hand.
The effectiveness of the recommender system is conditional on the slice granularity, and we showed that applying automatic transformations to fix this can be beneficial.

\newpage
\printbibliography

\newpage
\appendix
\section{Tier-specific annotations present in previous work}
\label{app:annotations}
\begin{table}[!htbp]
\caption{Supported annotations. The bottom row designates what type of statements the annotations can be placed on.}
\centering
\scriptsize{
\begin{tabular}{l l l l}
\hline
location-based & communication &  replication & failure handling  \\ \hline 
\textsc{@client}  & \textsc{@remoteCall}  	& \textsc{@local}		&\textsc{ @defineHandler} \\
\textsc{@server} & \textsc{@localCall}	& \textsc{@copy} 		& \textsc{@useHandler} \\
\textsc{@ui} 	& \textsc{@blocking} 	& \textsc{@replicated}	& 	\\
			& \textsc{@reply}		& \textsc{@observable}	&	 \\
			& \textsc{@remoteProcedure} &					& 	\\ \hline
block level & call or function level & declaration level & tier or call level \\ \hline
\end{tabular}
}
\label{tbl:stip:annotations}
\end{table}%

\section{Scenario-based evaluation}
\label{app:scenarioeval}
\begin{table}[!htbp]
\caption{Runtime results for the individual services of the Uni-corn app.}
\centering
\scriptsize{
\begin{tabularx}{0.85\textwidth}{c *{7}{Y}}
\hline
 & number of requests & failed requests & $\frac{\text{failed requests}}{\text{total requests}}$ & kb transferred & number of steps & $\frac{\text{successful steps}}{\text{total steps}}$  \\ \hline
\multicolumn{7}{c}{Tasks: view, add, update } \\
meteor & 2 & 0 & 0\,$\%$ & 1,083 & 3 & 100\,\% \\
hop.js & 5 & 0 & 0\,$\%$ & 20,8 & 3 & 100\,\%\\
Stip.js & 5 & 0 & 0\,$\%$ & 0,96 & 3 & 100\,\%\\ \hline

\multicolumn{7}{c}{Tasks offline: view, add, update} \\
meteor & 2 & 0 & 0\,$\%$ & 0,32 & 3 & 100\,\%\\
hop.js & 3 & 3 & 100\,$\%$ & 0 & 3 & 33\,\% \\
Stip.js & 0 & 0 & 0\,$\%$ & 0 & 3 & 100\,\%\\  \hline

\multicolumn{7}{c}{Meetings: view, add, update } \\
meteor & 2 & 0 & 0\,$\%$ & 1,185 & 3 & 100\,\% \\
hop.js & 5 & 0 & 0$\%$ & 10 & 3 & 100\,\%\\
Stip.js & 6 & 0 & 0\,\,$\%$ & 1,41 & 3 & 100\,\%\\  \hline

\multicolumn{7}{c}{Meetings offline: view, add, update} \\
meteor & 2 & 0 & 0\,$\%$ & 0,498 & 3 & 100\,\%\\
hop.js & 3 & 3 & 100\,$\%$ & 0 & 3 & 33\,\% \\
Stip.js & 0 & 0 & 0\,$\%$ & 0 & 3 & 100\,\%\\  \hline

\multicolumn{7}{c}{Schedule: view} \\
meteor & 0 & 0 & 0\,$\%$ & 0 & 1 & 100\,\% \\
hop.js & ? & ? & ?\,$\%$ & ? & ? & ?\,\%\\
Stip.js & 0 & 0 & 0\,$\%$ & 0 & 1 & 100\,\%\\  \hline

\multicolumn{7}{c}{Schedule offline: view} \\
meteor & 0 & 0 & 0\,$\%$ & 0 & 1 & 100\,\%\\
hop.js & ? & ? & ?\,$\%$ & ? & ? & ?\,\% \\
Stip.js & 0 & 0 & 0\,$\%$ & 0 & 1 & 100\,\%\\  \hline

\multicolumn{7}{c}{Charts: view} \\
meteor & 2 & 0 & 0\,$\%$ & 1,083 & 1 & 100\,\% \\
hop.js & 3 & 0 & 0\,$\%$ & 744 & 1 & 100\,\%\\
Stip.js & 0 & 0 & 0\,$\%$ & 0 & 1 & 100\,\%\\  \hline

\multicolumn{7}{c}{Charts offline: view} \\
meteor & 0 & 0 & 0\,$\%$ & 0 & 1 & 100\,\%\\
hop.js & 3 & 3 & 100\,$\%$ & 0 & 1 & 0\,\% \\
Stip.js & 0 & 0 & 0\,$\%$ & 0 & 1 & 100\,\%\\  \hline
\end{tabularx}
}
\label{tbl:eval:individualservices}
\end{table}%

\begin{table}
\caption{Individual steps of three scenarios for evaluating the Uni-corn application}
\centering
\scriptsize{
\begin{tabularx}{0.85\textwidth}{s c c c}
\hline
step & Scenario 1 & Scenario 2 & Scenario 3 \\ \hline
1 & View all tasks & View all meetings & \emph{Go offline}\\
2 & Add a new task & Add a new meeting & View all tasks\\
3 & Update the newly added task & Update the newly added meeting & Add a new task\\
4 & \emph{Go offline} & \emph{Go offline} & View all meetings\\
5 & Add a new task & Add a new meeting & Add a new meeting\\
6 & Update the newly added task &Update the newly added meeting & View the charts section\\
7 & View the charts section & View the calendar section &\emph{Go online}\\
8 &  \emph{Go online}& \emph{Go online} & View the charts section\\
9 & View all tasks & View all meetings& \\ \hline
\end{tabularx}}
\label{tbl:overviewscenarios}
\end{table}%

\end{document}